\documentclass{article}
\usepackage{amsfonts}
\usepackage{amssymb}
\usepackage{graphicx}
\usepackage{amsmath}

\numberwithin{equation}{section}

\input{tcilatex}

\begin{document}

\title{Optimization of Quantum Information Processing Maximizing the Mutual
Information\thanks{%
Originally published in: Radio Eng. Electron. Phys., \textbf{19} (9),
p.\thinspace 1349, 1973. [trans. from Radiotekhnika i Electronika, 1973, 
\textbf{19}, 9, 1839. 844]}}
\author{V\ P\ Belavkin \\
School of Mathematical Sciences, University of Nottingham, \\
Nottingham NG7 2RD \\
E-mail\textup{:} vpb@maths.nott.ac.uk  \and  R\ L\ Stratonovich}
\date{}
\maketitle

\begin{abstract}
A model of quantum noisy channel with input encoding by a classical random
vector is described. An equation of optimality is derived to determine a
complete set of wave functions describing quantum decodings based on
quasi-measurements maximizing the classical amount of transmitted
information. A solution of this equation is found for the Gaussian multimode
case with input Gaussian distribution. It is described by the overcomplete
family of coherent vectors describing an optimal quasimeasurement of the
canonical annihilation amplitudes in the output Hilbert space. It is found
that the optimal decoding in this case realizes the maximum amount $\mathsf{I%
}=\mathrm{Sp}\ln \big[1+S/(N+1)\big]$ of the classical information as
transmitted via the classical Gaussian channel with the effective noise
covariance matrix $N+I$. A physical realization of optimal quasi-measurement
based on an indirect (heterodyne) observation of the canonical operators is
suggested.
\end{abstract}

\section*{Introduction}

The use of homodyne or phase-sensitive receivers for the reception of a
coherently modulated quantum signal enables one to extract in the
quaiclassical limit only a half of the encoded information. The coherent
reception of quantum signals, based on simultaneous direct measurement of
amplitude and phase (or phase coordinates of each mode of the signal), is
forbidden due to the uncertainty relations of quantum theory. Therefore, it
is of interest to investigate the potential possibilities of decoding in
quantum communication channels based on the indirect, heterodyne methods of
observation. The theory of indirect\emph{\ }measurements of incompatible
observables $\left\{ b_{j}\right\} $ has been under the development starting
from the pioneering studies \cite{1,3,2} to meet the demands of the quantum
communication theory. Using a quantum reformulation of the statistical
inference theory for a Bayesian risk criteria it was recently shown how to
formulate optimal quantum information processings based on the indirect
measurements in the classical-quantum noisy channels.

The use of the concept of indirect measurement has made it possible to
indicate for a Gaussian classical-quantum communication channel the amount
of information 
\begin{equation*}
\mathsf{I}=\ln \left( 1+\frac{s}{n+1}\right) ,
\end{equation*}%
transmitted by each mode of coherently modulated signal in using a quantum
receiver with a large linear amplifier\footnote{%
This result was presented by R.L. Stratonovich at the symposium on
information theory (Dubna, June 15--25, 1969).} ($s,n$ are the energies of
the corresponding modes of the classical input signal and of the quantum
additive noise expressed in units of $h\nu $). The same amount of
information was deduced in~\cite{4}, where a method of its decoding was
indicated, leading to indirect measurement. The rate of information
transmission by the coherent signal was also determined in an earlier work~%
\cite{5}, where vectors of the $\left\{ \varphi _{\beta }\right\} $ coherent
states were used as the operators characterizing the coherent reception. The
problem of the realization of the coherent measurement and its optimality
was not considered. However the natural question -- whether the decoding
based on the quasi-measurement described by the coherent vectors \eqref{four}
is optimal, and if so,  in what conditions it is unique, has not been
resolved. 

In order to answer this question we derive the equation determining the
optimal decoding vectors $\left\{ \varphi _{\beta }\right\} $ according to
the criterion of the maximum amount of information and specify the optimal
indirect measurement realizing the corresponding quasi-measurement. 

\section{Decodings based on quasi-measurements}

\label{decod}

\emph{Quasi-measurement} of noncommuting operators $\left\{ b_{j}\right\} $
with values in a given (measurable) space $B\ni \beta $ are described by
operator-valued (positive, $\sigma $-additive) measures $\Pi (\mathrm{d}%
\beta )$ normalized to the identity operator $\hat{1}$ in the Hilbert space $%
\mathbb{H}$ of quantum states as $\Pi \left( B\right) =\hat{1}$ such that 
\begin{equation}
\hat{1}=\int \Pi (\mathrm{d}\beta ),\;\;\;\int \beta _{j}\Pi (\mathrm{d}%
\beta )=b_{j}.  \label{one}
\end{equation}%
If the resolution of identity in \eqref{one} is orthogonal such that $\Pi (%
\mathrm{d}\beta )\Pi (\mathrm{d}\beta ^{\prime })=\Pi (\mathrm{d}\beta \cap 
\mathrm{d}\beta ^{\prime })$, then it describes the usual (direct)
measurement of compatible observables in the spectral decomposition $%
b_{j}=\int \beta _{j}\Pi (\mathrm{d}\beta )$. In the general case, however,
the operators $\left\{ b_{j}\right\} $ do not have common orthogonal
spectral decomposition, and it can be said that the nonorthogonal expansion %
\eqref{one} describes their simultaneous quasi-measurement. The
probabilistic operator measure $\Pi (\mathrm{d}\beta )$ is an analog of
classical randomized decision rules $P(\mathrm{d}\beta |b)$.

The classical probability distribution $P(\mathrm{d}\beta )$ of the results
of quasi-measurements in the state specified by the statistical density
operator $\rho $ is determined by the Hilbert space trace $\mathrm{Tr}$ of
its product with the quantum probability measure $\Pi (\mathrm{d}\beta )$ in
the same way as in the case of direct measurements: 
\begin{equation}
P(\mathrm{d}\beta )=\mathrm{Tr}\,\Pi (\mathrm{d}\beta )\rho .  \label{two}
\end{equation}%
Below we shall specify the operator $\Pi (\mathrm{d}\beta )$ in the form 
\begin{equation}
\Pi (\mathrm{d}\beta )=\varphi _{\beta }\varphi _{\beta }^{\ast }\mu (%
\mathrm{d}\beta ),  \label{three}
\end{equation}%
explicitly taking account of its positive definiteness and $\sigma $%
-additiveness. Here $\{\varphi _{\beta },\beta \in B\}$ is a family of
non-Hermitian operators acting in general from a Hilbert space $\mathbb{F}$
into the original space $\mathbb{H}$ and, together with a positive measure $%
\mu (\mathrm{d}\beta )$ as the reference measure on the set $B$, satisfying
the condition of normalization \eqref{one} as the completeness condition for 
$\{\varphi _{\beta }\}$. In particular, if $\{\varphi _{\beta }\}$ are
operators acting from one-dimensional complex space $\mathbb{F}=\mathbb{C}$
into $\mathbb{H}$, i.e., are a set of wave functions $\varphi _{\beta }\in 
\mathbb{H}$ (in general nonorthogonal and nonnormalized: $\varphi _{\beta
}^{\ast }\varphi _{\beta ^{\prime }}\neq \delta _{\beta \beta ^{\prime }}$),
then the quasi-measurement specified by them is called elementary. We note
that any quasi-measurement can be realized by direct measurement of a
certain family of compatible observables $\left\{ \hat{\beta}_{j}\right\} $
in an expanded quantum system consisting of the original system and an
independent auxiliary system. This follows from the well-known fact that any
operator measure $\Pi (\mathrm{d}\beta )$ can be represented as partial
averaging $\Pi (\mathrm{d}\beta )=\mathrm{Tr}_{\mathbb{H}_{0}}\Theta (%
\mathrm{d}\beta )\rho _{0}$ of an orthoprojector measure $\Theta (\mathrm{d}%
\beta )$ given by common eigenvectors of the commuting operators $\hat{\beta}%
_{j}=\int \beta _{j}\Theta (\mathrm{d}\beta )$ acting in the expanded space $%
\mathbb{H}\otimes \mathbb{H}_{0}$, along with the state $\rho _{0}$ of the
auxiliary system described by a Hilbert space $\mathbb{H}_{0}$. Such a
realization of quasi-measurement is called \emph{indirect measurement}. 

As an example we consider an elementary quasi-measurement specified by the
coherent vectors 
\begin{equation}
\varphi _{\alpha }=\exp \left\{ -\frac{1}{2}\alpha ^{\dagger }\alpha
+a^{\dagger }\alpha \right\} |0\rangle =|\alpha \rangle .  \label{four}
\end{equation}%
It can be realized by direct simultaneous measurement of the complex
amplitudes 
\begin{equation}
\hat{\alpha}=a+a_{0}^{+}=\left( \hat{\alpha}_{j}\right) ,\;\;\;\;j=1,\ldots
,r,  \label{five}
\end{equation}%
where $\alpha =\left( \alpha _{j}\right) $ and  $a=(a_{j})$ are columns of
the amplitudes $\alpha _{j}\in \mathbb{C}$\ and the annihilation operators $%
a_{j}$ operating into the original Hilbert space $\mathbb{H}$, $\alpha
^{\dagger }$ and $a^{\dagger }$ are the adjoint raws, while $a_{0}^{+}$
denotes the column $(a_{0j}^{+})$ of creation operators $a_{0j}^{+}$ of an
auxiliary (heterodyning) system in the vacuum state $|0\rangle _{0}\in 
\mathbb{H}_{0}$. For this purpose it is sufficient to note that the partial
averaging over the vacuum state $|0\rangle _{0}$ amounts to the
determination of the projections $\varphi _{\alpha }$ of the eigenvectors $%
\psi _{\alpha }\in \mathbb{H}\otimes \mathbb{H}_{0}$ of commuting operators $%
\alpha =a+a_{0}^{+}$ in the original space $\mathbb{H}$. One can easily
verify that the generalized vectors 
\begin{equation}
\psi _{\alpha }=\exp \big\{-(a-\alpha )^{\dagger }a_{0}^{+}\big\}\,|0\rangle
_{0}\otimes |\alpha \rangle   \label{six}
\end{equation}%
satisfy the equations $\hat{\alpha}\psi _{\alpha }=\alpha \psi _{\alpha }$, $%
\hat{\alpha}^{\ast }\psi _{\alpha }=\alpha ^{\ast }\psi _{\alpha }$ and form
a complete orthonormal set with the measure 
\begin{equation*}
\mathrm{d}\mu (\alpha )=\prod_{j=1}^{r}\tfrac{1}{\pi }\mathrm{d}\func{Re}%
\alpha _{j}\mathrm{d}\func{Im}\alpha _{j}.
\end{equation*}%
Multiplying these vectors from the left by ${}_{0}\langle 0|$ we find that
the projections $\varphi _{\alpha }\in \mathbb{H}$ are actually the coherent
vectors $|\alpha \rangle $ and, together with the measure $\mu (\mathrm{d}%
\alpha )=\mathrm{d}\mu (\alpha )$, determine ideal measurement corresponding
to the indirect measurement of noncommuting $a_{j}$.

The question -- whether the decoding based on the quasi-measurement
described by the coherent vectors \eqref{four} is optimum and if so in what
conditions -- is of interest. In order to answer this question we are going
to derive the equation determining the operators $\varphi _{\beta }$, which
are optimal according to the criterion of the maximum amount of mutual
information and specify the optimum quasi-measurement.

We shall assume that the family of statistical operators $\big\{\rho
(\vartheta )\big\}$ defining the state of the quantum communication channel
is given as a function of the random information parameters $\vartheta
=(\vartheta _{j})$ having the distribution $P(\mathrm{d}\vartheta )$. The
amount of Shannon information of the parameters $\vartheta $ and the results
of quasi-measurements $\beta $ is determined by the usual formula 
\begin{equation*}
\mathsf{I}_{\beta ,\vartheta }=\iint p(\beta |\vartheta )\ln \frac{p(\beta
|\vartheta )}{p\left( \beta \right) }P(\mathrm{d}\vartheta )\mu (\mathrm{d}%
\beta ),
\end{equation*}%
where according to \eqref{two}, \eqref{three} the density $p(\beta
|\vartheta )$ has the form 
\begin{equation}
p(\beta |\vartheta )=\mathrm{Tr}\,\varphi _{\beta }\varphi _{\beta }^{\ast
}\rho (\vartheta )=\mathrm{Tr}\,\varphi _{\beta }^{\ast }\rho (\vartheta
)\varphi _{\beta },  \label{seven}
\end{equation}%
$p\left( \beta \right) =\int p\left( \beta |\vartheta \right) P(\mathrm{d}%
\vartheta )$, satisfying the normalization condition 
\begin{equation*}
\int p(\beta |\vartheta )\mu (\mathrm{d}\beta )=1,\;\;\int p\left( \beta
\right) \mu (\mathrm{d}\beta )=1.
\end{equation*}%
Using the method of Lagrangian multipliers we set up a function 
\begin{equation*}
\int \left( \int i(\beta ,\vartheta )\mathrm{Tr}\,\varphi _{\beta }^{\ast
}\rho (\vartheta )\varphi _{\beta }P(\mathrm{d}\vartheta )-\mathrm{Tr}%
\,\varphi _{\beta }^{\ast }\hat{\lambda}\varphi _{\beta }\right) \,\mu (%
\mathrm{d}\beta ),
\end{equation*}%
where $i(\beta ,\vartheta )=\ln \left[ p(\beta |\vartheta )\big/\int p(\beta
|\vartheta )P(\mathrm{d}\vartheta )\right] $, and $\hat{\lambda}$ is an
operator determined from the condition of completeness $\int \varphi _{\beta
}\varphi _{\beta }^{\ast }\mu (\mathrm{d}\beta )=\hat{1}$. Varying it over $%
\varphi _{\beta }^{\ast }$ we obtain the equation determining optimum $%
\varphi _{\beta }$: 
\begin{equation}
\left( I\left( \beta \right) -\hat{\lambda}\right) \varphi _{\beta }=0,
\label{eight}
\end{equation}%
where $I\left( \beta \right) =\int i(\beta ,\vartheta )\rho (\vartheta )P(%
\mathrm{d}\vartheta )$ and $\hat{\lambda}=\int I\left( \beta \right) \varphi
_{\beta }\varphi _{\beta }^{\ast }\,\mu (\mathrm{d}\beta )$.

In the derivation of this equation we have made use of the fact that%
\begin{equation*}
\int p(\beta |\vartheta )\delta i(\beta ,\vartheta )(\beta ,\vartheta )P(%
\mathrm{d}\vartheta )=\int p(\beta |\vartheta )\delta \varphi ^{\ast }\left( 
\frac{\rho (\vartheta )}{p(\beta |\vartheta )}-\frac{\rho }{p(\beta )}%
\right) \varphi _{\beta }P(\mathrm{d}\vartheta )=0,
\end{equation*}%
where $\rho =\int \rho (\vartheta )P(\mathrm{d}\vartheta )$. Multiplying
equation~\eqref{eight} from the right by $\varphi _{\beta }^{\ast }$ and
integrating with the measure $\mu (\mathrm{d}\beta )$ we determine the
operator $\hat{\lambda}$: 
\begin{equation*}
\hat{\lambda}=\iint \rho (\vartheta )i(\beta ,\vartheta )\varphi _{\beta
}\varphi _{\beta }^{\ast }\mu (\mathrm{d}\beta )P(\mathrm{d}\vartheta ),
\end{equation*}%
the trace of which $\mathrm{Tr}\,\hat{\lambda}$ gives the maximum amount of
information. And, finally, eliminating the operator from Equation.~%
\eqref{eight} and multiplying it from the left by $\hat{\lambda}$, we
rewrite this equation in the following equivalent (because of the condition
of completeness $\int \varphi _{\alpha }\varphi _{\alpha }^{\ast }\mu (%
\mathrm{d}\alpha )=\hat{1}$) form: 
\begin{equation}
\int \varphi _{\alpha }^{\ast }\rho (\vartheta )\left( \varphi _{\beta
}i(\beta ,\vartheta )-\int \varphi _{\beta ^{\prime }}\varphi _{\beta
^{\prime }}^{\ast }\varphi _{\beta }\mathsf{I}(\beta ^{\prime },\vartheta )u(%
\mathrm{d}\beta ^{\prime })\right) P(\mathrm{d}\vartheta )=0.  \label{nine}
\end{equation}%
The equation thus obtained is a complex nonlinear equation in $\varphi
_{\beta }$ and it is not possible to solve it explicitly in the general
case. However, as will be shown in the next section, in the Gaussian case
the coherent vectors~\eqref{four} are optimum operators $\varphi _{\beta }$.

\section{Optimal decoding in quantum Gaussian channel}

\label{BG channel}

Let the received signal $b$ be a superposition $b=\vartheta +a$ of a complex
vector $\vartheta =(\vartheta _{j},j=1,\ldots ,r)$ and an independent
Gaussian Boson noise $a=(a_{j})$ ($a_{j}$ are annihilation operators of the
investigated modes: $[a_{j},a_{k}^{+}]=\delta _{jk}$, having zero
mathematical expectation $\langle a_{j}\rangle =0$ and specified average
number of quanta $\langle a_{j}^{+}a_{j}\rangle =n_{j}$). In Glauber's
representation~\cite{6} the statistical operator of this noise has the form 
\begin{equation}
\rho =\int |\alpha \rangle \langle \alpha |\,|N|^{-1}\exp \{-\alpha
^{\dagger }N^{-1}\alpha \}\,\mathrm{d}\mu (\alpha ),  \label{ten}
\end{equation}%
where $|\alpha \rangle $ is the coherent vectors~\eqref{four}; $N$ is a
matrix whose eigenvalues are $n_{j}$, and $|N|=\det N=n_{1}\cdots n_{r}$. In
formula~\eqref{ten} the matrix form of writing scalar products is used, $%
\alpha ^{\dagger }N^{-1}\alpha =\sum_{j,k}\alpha _{j}^{\ast
}(N^{-1})_{jk}\alpha _{k}$ , similar as $a^{\dagger }\alpha
=\sum_{j}a_{j}^{+}\alpha _{j}$ in~\eqref{four}. The signal $b$ after passing
through the indicated linear channel is described by a family of operators $%
\big\{\rho (\vartheta )\big\}$ of the form 
\begin{equation}
\rho (\vartheta )=\int |\alpha \rangle \langle \alpha |p(\alpha |\vartheta
)\,\mathrm{d}\mu (\alpha ),  \label{eleven}
\end{equation}%
where $p(\alpha |\vartheta )=|N|^{-1}\exp \big\{-(\alpha -\vartheta
)^{\dagger }N^{-1}(\alpha -\vartheta )\big\}$. The probability distribution $%
P(\mathrm{d}\vartheta )$ of the transmitted signal $\vartheta $, chosen from
the condition of maximum entropy and finiteness of the energy $\langle
\vartheta ^{\dagger }S^{-1}\vartheta \rangle \leq $ ($S$ is the given
correlation matrix $\langle \vartheta _{j}^{\ast }\vartheta _{k}\rangle $),
also has a Gaussian form: 
\begin{equation}
P(\mathrm{d}\vartheta )=|S|^{-1}\exp \{-\vartheta ^{\dagger }S^{-1}\vartheta
\}\,\mathrm{d}\mu (\vartheta ),\quad \mathrm{d}\mu (\vartheta
)=\prod_{j=1}^{r}\tfrac{1}{\pi }\mathrm{d}\func{Re}\vartheta _{j}\,\mathrm{d}%
\func{Im}\vartheta _{j}.  \label{twelve}
\end{equation}

In the case under investigation it is natural to expect that the results of
optimum quasi-measurement of the Gaussian observable $b$ also have a
Gaussian distribution; among the Gaussian operators $\varphi _{\beta }$,
determining this distribution, it is natural to separate out the operators
characterized by the minimum uncertainty relation $\left\langle (b-\beta
)_{j}^{\dagger }(b-\beta )_{k}\right\rangle =\delta _{jk}$. Such operators
are coherent vectors.

Let us verify if vectors $\varphi _{\beta }$ of form~\eqref{four} satisfy
Equation.~\eqref{nine} in the Gaussian case. Substituting~\eqref{four}, %
\eqref{eleven} into \eqref{seven} and taking into account that%
\begin{equation*}
\langle \beta |\alpha \rangle =\exp \big\{\beta ^{\dagger }\alpha -(\alpha
^{\dagger }\alpha +\beta ^{\dagger }\beta )/2\big\}
\end{equation*}%
we can find the conditional density 
\begin{align}
\lefteqn{p(\beta |\vartheta )=\int \langle \beta |\alpha \rangle \langle
\alpha |\beta \rangle p(\alpha |\vartheta )\,\mathrm{d}\mu (\alpha )}  \notag
\label{thirteen} \\
& =|N+I|^{-1}\exp \big\{-(\beta -\vartheta )^{\dagger }(N+I)^{-1}(\beta
-\vartheta )\big\},
\end{align}%
and also the function $i(\beta ,\vartheta )$ occurring in Equation.~%
\eqref{nine}: 
\begin{equation}
i(\beta ,\vartheta )=\ln \left\vert I+S(N+I)^{-1}\right\vert +\vartheta
^{\dagger }S^{-1}\vartheta -(\vartheta -A\beta )^{\dagger }G(\vartheta
-A\beta ),  \label{fourteen}
\end{equation}%
where $A=S(S+N+I)^{-1}$, and $G=S^{-1}+(N+I)^{-1}$. In~\eqref{fourteen} only
the last term is significant for Equation.~\eqref{nine}: the first two terms
drop out after substitution into this equation. Substituting the last term $%
(A\beta -\vartheta )^{\dagger }G(A\beta -\vartheta )\equiv c(A\beta
,\vartheta )$ of function~\eqref{fourteen} into Equation.~\eqref{nine} and
using representation~\eqref{eleven} of the operator $\rho (\vartheta )$, we
verify that the coherent vector $\varphi _{\beta }=|\beta \rangle $ is a
solution of Equation.~\eqref{nine}. Thus the solution of the formulated
problem is reduced, as also in the optimization according to the mean square
criterion, to the verification of the identity 
\begin{align}
& \iint \langle \alpha |\alpha ^{\prime }\rangle \langle \alpha ^{\prime
}|\beta \rangle \left( c(A\beta ,\vartheta )-\int \frac{\langle \alpha
^{\prime }|\beta ^{\prime }\rangle \langle \beta ^{\prime }|\beta \rangle }{%
\alpha ^{\prime }|\beta \rangle }c(A\beta ^{\prime },\vartheta )\,\mathrm{d}%
\mu (\beta ^{\prime })\right)   \notag  \label{fifteen} \\
& \times p(\alpha ^{\prime }|\vartheta )\,\mathrm{d}\mu (\alpha ^{\prime })P(%
\mathrm{d}\vartheta )=0,
\end{align}%
where $p(\alpha |\vartheta )$, $P(\mathrm{d}\vartheta )$ are defined in~%
\eqref{eleven}, \eqref{twelve}.

Below we shall require the following formulas of integration in the complex $%
r$-dimensional space $\mathbb{C}^{r}\ni z$: 
\begin{align}
& \int \exp \big\{-(z-\alpha )^{\dagger }Q(z-\beta )\big\}|Q|\,\mathrm{d}\mu
(z)=1,  \notag  \label{sixteen} \\
& \int z\exp \big\{-(z-\alpha )^{\dagger }Q(z-\beta )\big\}|Q|\,\mathrm{d}%
\mu (z)=\beta ,  \notag \\
& \int z^{\ast }\exp \big\{-(z-\alpha )^{\dagger }Q(z-\beta )\big\}|Q|\,%
\mathrm{d}\mu (z)=\alpha ^{\ast }, \\
& \int (z-\alpha )^{\dagger }H(z-\beta )\exp \big\{-(z-\alpha )^{\dagger
}Q(z-\beta )\big\}|Q|\,\mathrm{d}\mu (z)=\mathrm{Sp}\,Q^{-1}H,
\end{align}%
all positive-definite $(r\times r)$-matrices $Q,H$, where $\mathrm{Sp}$ is
the trace in the space $\mathbb{C}^{r}$ where all vectors $\alpha ,\beta \in 
\mathbb{C}^{r}$ are taking values. Taking into account that%
\begin{equation*}
\langle \alpha ^{\prime }|\beta \rangle \langle \beta ^{\prime }|\beta
\rangle \big(\langle \alpha ^{\prime }|\beta \rangle \big)^{-1}=\exp \big\{%
-(\beta ^{\prime }-\alpha ^{\prime })^{\dagger }(\beta ^{\prime }-\beta )%
\big\},
\end{equation*}%
in the identity~\eqref{fifteen} we can carry out the integration over after
making the substitution $A\beta ^{\prime }=z$ and making use of formula~%
\eqref{sixteen} for $Q=(AA^{\dagger })^{-1},H=G$. As a result the expression
in the parentheses in~\eqref{fifteen} becomes $(\beta -\alpha ^{\prime
})^{\dagger }A^{\dagger }G(A\beta -\vartheta )\mathrm{Sp}\,AA^{\dagger }G$.
The integration of the last expression over $\alpha ^{\prime }$ with the
density 
\begin{align*}
\lefteqn{\langle \alpha |\alpha ^{\prime }\rangle \langle \alpha ^{\prime
}|\beta \rangle p(\alpha ^{\prime }|\vartheta )} \\
& =\langle \alpha |\beta \rangle |N+I|^{-1}|\bar{Q}|\exp \big\{-(\vartheta
-\alpha )^{\dagger }(N+I)^{-1}(\vartheta -\beta )-(\alpha ^{\prime }-\bar{%
\alpha})^{\dagger }\bar{Q}(\alpha ^{\prime }-\beta )\big\}
\end{align*}%
amounts to the use of the first and the third formulas of integration~%
\eqref{sixteen}, where for $Q,\alpha ,\beta $ we should take $\bar{Q}\equiv
N^{-1}+I$, $\,\bar{\alpha}\equiv (N+I)^{-1}(\vartheta +N\alpha )$, $\,\beta
\equiv (N+I)^{-1}(\vartheta +N\beta )$. This gives 
\begin{align}
& \Big[\big(\beta -(N+I)^{-1}(\vartheta +N\alpha )\big)^{\dagger }A^{\dagger
}G(A\beta -\vartheta )-\mathrm{Sp}\,AA^{\dagger }G\Big]\,\langle \alpha
|\beta \rangle |  \notag  \label{seventeen} \\
& \times |N+I|^{-1}\exp \big\{-(\vartheta -\alpha )^{\dagger
}(N+I)^{-1}(\vartheta -\beta )\big\}.
\end{align}%
In order to complete the verification of identity~\eqref{fifteen} it only
remains to carry out averaging with the distribution $P(\mathrm{d}\vartheta )
$. Rewriting the expression in the square brackets in~\eqref{seventeen} in
the form 
\begin{equation*}
\big(\vartheta -(N+I)\beta +N\alpha \big)^{\dagger }(N+I)^{-1}A^{\dagger
}G(\vartheta -A\beta )-\mathrm{Sp}\,AA^{\dagger }G
\end{equation*}%
and integrating it with respect to $\vartheta $ with the density 
\begin{align*}
& |N+I|^{-1}|S|^{-1}\exp \big\{-(\vartheta -\alpha )^{\dagger
}(N+I)^{-1}(\vartheta -\beta )-\vartheta ^{\dagger }S^{-1}\vartheta \big\} \\
& =|G|\,|S+N+I|e^{-(\vartheta -A\alpha )^{\dagger }G(\vartheta -A\beta )}
\end{align*}%
according to the formulas~\eqref{sixteen} for $Q=G,H=(N+I)^{-1}A^{\dagger }G$
we obtain 
\begin{equation*}
|S+N+I|^{-1}\mathrm{Sp}\,\big(G^{-1}(N+I)^{-1}A^{\dagger }-AA^{\dagger }\big)%
G.
\end{equation*}

This expression coincides with the left-hand side of~\eqref{fifteen} with an
accuracy up to a nonzero factor $\langle \alpha |\beta \rangle $. Due to the
identity 
\begin{equation*}
G^{-1}=\big(S^{-1}(N+I)^{-1}\big)^{-1}=S(S+N+I)^{-1}(N+I)=A(N+I),
\end{equation*}%
as a consequence of which%
\begin{equation*}
\mathrm{Sp}\,\big(G^{-1}(N+I)^{-1}A^{\dagger }-AA^{\dagger }\big)G=\mathrm{Sp%
}\,(AA^{\dagger }-AA^{\dagger })G=0,
\end{equation*}%
the validity of the verified equation is follows straightforward.

Thus in a linear Gaussian Bosonic channel the optimum decoding is given by
coherent vectors of the form~\eqref{four}. The ideal quasi-measurement,
defined by the overcomplete nonorthogonal set $\big\{|\beta \rangle \big\}$
of coherent vectors, is realized by the measurement of complex amplitudes 
\begin{equation}
\hat{\beta}=b+a_{0}^{+}=\vartheta +a+a_{0}^{+}=\vartheta +\alpha ,
\label{eighteen}
\end{equation}%
in the expanded space these amplitudes are of a complete orthonormal set of
eigenvectors. The measurement of simultaneous observables~\eqref{eighteen}
is a linear indirect measurement of noncommuting $b=(b_{j})$ investigated in~%
\cite{7}. The maximum amount of information decoded by the optimal ideal
quasi-measurement is easily found by averaging the function~\eqref{fourteen}:%
\begin{equation*}
\mathsf{I}=\ln |I+S(N+I)^{-1}|=\mathrm{Sp}\,\ln \big(I+S(N+I)^{-1}\big).
\end{equation*}%
In the one-dimensional case the obtained amount of information coincides
with that shown in~\cite{4,5}.

In conclusion, we note that if a Gaussian stationary signal transmitted
along the channel is mixed with stable thermal noise of temperature $\theta
=kT$, then the rate information transmission in optimum decoding is
determined by the information%
\begin{equation*}
\ln \big(1+s_{\nu }/(n_{\nu }+1)\big)
\end{equation*}%
transmitted by each mode summed over the operating frequency range $N\ni \nu 
$: 
\begin{equation}
\mathsf{I}=\int_{N}\ln \big(1+s_{\nu }(n_{\nu }+1)^{-1}\big)\,\mathrm{d}\nu
=\int_{N}\big(1+s_{\nu }(1-e^{-h\nu /\theta })\big)\,\mathrm{d}\nu .
\label{nineteen}
\end{equation}%
Here $s_{\nu }$ is the spectral intensity of the transmitted signal
expressed in units of $h\nu $, and $n_{\nu }=(e^{h\nu /\theta }-1)^{-1}$ is
the spectral intensity of the noise. In the classical limit $h\nu /\theta
\ll 1$ the rate of information transmission goes over into the corresponding
classical expression $\mathsf{I}=\int \ln (1+\sigma _{\nu }^{2}/\theta )%
\mathrm{d}\nu $, where $\sigma _{\nu }^{2}=s_{\nu }h\nu $. In the opposite
case $h\nu /\theta \gg 1$ of low temperatures the transmission rate $\mathsf{%
I}=\int \ln (1+s_{\nu })\mathrm{d}\nu $ remains finite in contrast to the
classical case. It is not difficult to see that the quantum corrections to~%
\eqref{nineteen} are significant only for weak (at the receiving end of the
channel) signals $s_{\nu }\ll 1$, for which formula~\eqref{nineteen} can be
written in the form $\mathsf{I}\approx \int (1-e^{-h\nu /\theta })s_{\nu }%
\mathrm{d}\nu $.

\end{document}